\documentclass[twocolumn,showpacs,preprintnumbers,amsmath,amssymb]{revtex4}

\usepackage{graphicx}
\usepackage{dcolumn}
\usepackage{bm}
\usepackage[latin1]{inputenc}     

\begin{document}

\preprint{}

\title{Realistic Picture of 2D Harmonic Oscillator Coherent States}

\author{Michel Gondran}
 \email{michel.gondran@chello.fr}
\affiliation{EDF, Research and Development, 1 av. du Général de Gaulle, 92140 Clamart, France.}%

\date{\today}

\begin{abstract}
We show that a 2D harmonic oscillator coherent state is a soliton
which has the same evolution as a spinning top: the center of mass
follows a classical trajectory and the particle rotates around its
center of mass in the same direction as its spin with the harmonic
oscillator frequency.
\end{abstract}


\maketitle

\section{Introduction}

The harmonic oscillator coherent states introduced in 1926 by
Schrödinger~\cite{Schrodinger_26} have become very important in
quantum optics due to Glauber~\cite{Glauber_65} in 1965, who
deduced the quantum theory of optical coherence from these
coherent states. Note its three most important properties: 1. it
describes a nondispersive wave packet, 2. its center follows a
classical trajectory, 3. the Heisenberg inequalities are
equalities.

\bigskip

We show in this paper that a coherent state of 2D harmonic
oscillator can be considered as a solid which rotates around its
center of mass in the same direction as its spin with an angular
velocity $\omega$, while the center of mass follows the classical
trajectory of a harmonic oscillator of frequency $\omega$.

The proof presented in this article will use a new definition of
the Schrödinger current that shall be recalled here.

\section{The Schrödinger current}

In 1928 Gordon~\cite{Gordon} showed that the Dirac current can be
subdivided into a convection current and a spin-dependent current.
In the Pauli non relativistic approximation, this spin-dependent
current can be written~\cite{Gordon}:
\begin{equation}\label{courantgordon}
    {\textbf{ J}_{Pauli-spin}}=\frac{\hbar}{2m_{e}}\nabla \times (\Psi^* \sigma
    \Psi).
\end{equation}
Let's suppose now that the particle is in a spin eigenstate, i.e.
that the Pauli spinor can be written  $ \Psi(\textbf{r},t)=
\psi(\textbf{r},t) \chi$ where $\chi$ is a constant spinor such as
$\chi^{*} \chi=1$. Holland~\cite{Holland_99} shows that the Pauli
spin-dependent current becomes the Schrödinger spin-dependent
current
\begin{equation}\label{courantspin}
    { \textbf{J}_{Sch-spin}}=\frac{\hbar}{2 m_e} \nabla \rho
\times \textbf{u}
\end{equation}
where $\psi= \sqrt{\rho} e^{i \frac{S}{\hbar}}$ and where
$\textbf{u}=\chi^* \sigma \chi$ is the spin vector. So that, to
obtain a good approximation of the Dirac current, it is necessary
to add to the classical convection current ${\bf
J_{Sch-conv}}=\frac{i \hbar}{2m_{e}}(\psi \nabla\psi^* - \psi^*
\nabla\psi)=\rho \frac{\nabla S}{m_e}$ the spin-dependent
current~(\ref{courantspin}). Then the new Schrödinger current
becomes \cite{Holland_99}:
\begin{equation}\label{courantSch}
    { \textbf{J}_{Sch}}=\rho
  \frac{\nabla S}{m_e} + \frac{\hbar}{2 m_e} \nabla \rho
\times \textbf{u}.
\end{equation}

It is possible to verify from the ground state of the hydrogen
atom that this new definition is necessary. The classical
Schrödinger convection current of the eigenfunction $\psi_{100}$
(ground state) is zero because $\psi_{100}$ is real. Therefore, in
the ground state of the hydrogen atom $1s_{\frac{1}{2}}$, the
Dirac current is equal to ~\cite{Gondran_2003,Colijn_2003}
\begin{equation}\label{courantdirac1s}
 \textbf{J}_{Dirac1s_{\frac{1}{2}}}=\rho \alpha c \sin \theta {\bf
u_{\varphi}}.
\end{equation}
This is exactly equal to the value given by the Schrödinger
spin-dependent current~(\ref{courantspin}).

Finally, de Struyve, De Baere, De Neve and De
Weirdt~\cite{Struyve} have proved that the
formula~(\ref{courantSch}) is also necessary for the bosons of
spin $1$.

\section{Realistic picture of the 2D harmonic oscillator coherent states}

In the case of the 2D harmonic oscillator,
$V(\textbf{x})=\frac{1}{2}m \omega^{2}(x^{2}+y^2)$, the coherent
states are build~\cite{CohenTannoudji} on the initial wave
function $\Psi_{0}(\textbf{x})=\left( 2\pi \sigma
_{\hbar}^{2}\right) ^{-\frac{1}{2}}e^{-\frac{ (
\textbf{x}-\xi_{0}) ^{2}}{4\sigma _{\hbar}^{2}}+i \frac{m
\textbf{v}_0 . \textbf{x}}{\hbar} }$ where $\xi_0$ and
$\textbf{v}_0$ are independent data from $\hbar$ and where
$\sigma_\hbar =\sqrt{\frac{\hbar}{2 m \omega}}$.

The wave function $\Psi (\textbf{x},t)$, solution of the
Schrödinger equation is then the coherent
state~\cite{CohenTannoudji}:
\begin{equation}
\Psi(\textbf{x},t)=\left( 2\pi \sigma _{\hbar}^{2}\right)
^{-\frac{1}{2}}
  e^{-\frac{( \textbf{x}-\xi(t)) ^{2}}{4\sigma
_{\hbar}^{2}}+i \frac{m \textbf{v}(t). \textbf{x} -g(t)}{\hbar} }
\end{equation}
where $\xi(t)= \xi_0 \cos(\omega t)
+\frac{\textbf{v}_0}{\omega}\sin(\omega t)$ and $\textbf{v}(t)=
\textbf{v}_0 \cos(\omega t) -\xi_0 \omega \sin(\omega t)$
respectively correspond to position and velocity of a classical
particle in a potential $V(\textbf{x})=\frac{1}{2} m \omega^{2}
(x^{2}+y^2)$: $\xi_0$ and $\textbf{v}_0$ are the initial position
and velocity and $ g(t)=\int _0 ^t ( \hbar \omega + \frac{1}{2} m
\textbf{v}^{2}(s) - \frac{1}{2} m \omega^{2} \xi^{2}(s)) ds$.

The energy $E(\textbf{x},t)$ defined by equation $ i \hbar
\frac{\partial \Psi}{\partial t}(\textbf{x},t)=
E(\textbf{x},t)\Psi(\textbf{x},t) $ is equal to ~$g'(t) - m
\frac{d\textbf{v}(t)}{dt} . \textbf{x}+ i \hbar \frac{\textbf{x} -
\xi(t)}{2 \sigma^2_\hbar}. \textbf{v}(t)$. On the trajectory
$\xi(t)$, the energy is constant and equal to
\begin{eqnarray}\label{eq:energie}
E(\xi(t),t) &=&  \hbar \omega +(\frac{1}{2} m \textbf{v}^{2}(t) + \frac{1}{2} m \omega^{2} \xi^{2}(t))\nonumber\\%
            &=& \hbar \omega +(\frac{1}{2} m \textbf{v}_{0}^{2} + \frac{1}{2} m \omega^{2} \xi_0^{2}).%
\end{eqnarray}
Then, the coherent state is a state with a constant energy on the
trajectory $\xi(t)$.

In the Schrödinger approximation (constant spin orientation), the
velocity field $\textbf{v}(\mathbf{x},t)$ of the 2D harmonic
oscillator is equal to (cf.~(\ref{courantSch})):
\begin{eqnarray}\label{eq:eqvitessemqosc}
\textbf{v}(\mathbf{x},t) &=& \frac {1}{\rho}\textbf J_{Sch}(\textbf{x},t)%
                          = \frac{\nabla S(\mathbf{x},t)}{m} + \frac{\hbar\nabla\rho(\mathbf{x},t)}{2 m \rho(\mathbf{x},t)}\times\textbf{k}\nonumber\\%
                         &=& \textbf{v}(t) + \Omega \times (\textbf{x} -\xi(t))
\end{eqnarray}
\bigskip
where $\textbf{k}$ is the spin vector and where $\Omega = \omega
\textbf{k}$.

$\textbf{v}(\mathbf{x},t)$ can be interpreted as the velocity of a
solid. Its center of mass follows a classical trajectory. The
solid rotates around its center of mass in the same direction as
its spin with an angular velocity $\omega$.

We have found for the coherent states of the 2D harmonic
oscillator a classical geometric picture. We verify that this
picture corresponds to a spread particle which satisfies the
Heisenberg equalities
\begin{equation}\label{eq:heisenberg}
\Delta x . \Delta p_{x}=\frac{\hbar}{2};~~~~\Delta ~y . \Delta
p_{y}=\frac{\hbar}{2}
\end{equation}
and has the energy~(\ref{eq:energie}). Indeed, we have:
\begin{eqnarray*}
(\Delta x)^2 &=& < (x- \xi_{x}(t))^2>\\
             &=& \int (x- \xi_{x}(t))^2 \left( 2\pi \sigma _{\hbar}^{2}\right) ^{-\frac{1}{2}} e^{-\frac{(x- \xi_{x}(t) ) ^{2}}{4\sigma_{\hbar}^{2}}}dx%
              =\sigma_\hbar^2,
\end{eqnarray*}
\begin{eqnarray*}
(\Delta p_x)^2&=&<( m \textbf{v}_x(\textbf{x},t) - m \textbf{v}_x(t))^2>\\%
              &=&<m^2 \omega^2 (y- \xi_y(t))^2>= m^2 \omega^2 \sigma_\hbar^2,
\end{eqnarray*}
and
\begin{eqnarray*}
E &=& <\frac{1}{2} m \textbf{v}^2(\textbf{x},t) + \frac{1}{2} m \omega^2 \textbf{x}^2>\\%
  &=& <\frac{1}{2} m \textbf{v}^2(t)>+<\frac{1}{2} m \omega^2(\textbf{x}-\xi(t))^2>\\%
   && + <\frac{1}{2} m \omega^2 \xi(t)^2>+<\frac{1}{2} m \omega^2 (\textbf{x}-\xi(t))^2>\\%
  &=& \frac{1}{2} m \textbf{v}_{0}^{2} + \frac{1}{2} m \omega^{2} \xi_0^{2} + \hbar \omega.%
\end{eqnarray*}
Then, it is natural to assume, in this case, that the wave
function represents a spread particle and its square the particle
density.

The 2D harmonic oscillator ground state corresponds to $\xi_0=0$
and $\textbf{v}_0=0$. It can be represented by a disk of density
$\rho(x,y)= \left( 2\pi \sigma _{\hbar}^{2}\right)^{-1} e^{-\frac{
x^{2}+y^2}{2\sigma _{\hbar}^{2}}}$ which spins with an angular
velocity $\omega$.

\section{Conclusion}

We have proposed for the 2D oscillator harmonic wavefunction the
picture of a spread particle (soliton). However, we have still not
found such a simple picture for the 3D harmonic oscillator and the
hydrogen atom.

The hydrogen atom ground state, in the Schrödinger approximation,
is a coherent state as the harmonic oscillator ground state: the
wave packet center is also immobile ($\xi(t)=0$) and the velocity
is equal to $ \textbf{v}^{\hbar}(\mathbf{x},t) = \alpha c \sin
\theta {\bf u_{\varphi}}=\Omega\times\frac{\textbf{r}}{r}$ with
$\Omega = \alpha c \textbf{k}$.

Therefore, the picture of a spread particle of density $\rho(r)=
\frac{1}{\pi a^3_0}\ e^{-2\frac{r}{a_0}}$, where each point
$(r,\theta)$ rotates around the spin vector with a constant
velocity $\alpha c \sin\theta$, satisfies Heisenberg equalities $
\Delta r \Delta p= \sqrt{2} \hbar$ but not the energy
$E=-\frac{1}{2} m \alpha^2 c^2$.


\begin{thebibliography}{99}



\bibitem{Schrodinger_26}
E. Schrödinger, Naturwissenschaften \textbf{14}, 664 (1926).

\bibitem{Glauber_65}
R. J. Glauber, in \textit{Quantum Optics and Electronics}, edited by C. deWitt, A. Blandin and C. Cohen-Tanoudji (Gordon and Breach, New York, 1965).%

\bibitem{CohenTannoudji}
C. Cohen-Tannoudji, B.Diu, and F. Laloë, \textit{Quantum Mechanics} (Wiley, New York, 1977).%

\bibitem{Gordon}
W. Gordon, Z. Phys. {\bf 48}, 11 (1928); Z. Phys. {\bf 50}, 630 (1928).%

\bibitem{Holland_99}
P. R. Holland, Phys. Rev. A \textbf{60}, 6 (1999).

\bibitem{Gondran_2003}
M. Gondran and A. Gondran, quant-ph/0304055 (2003).%

\bibitem{Colijn_2003}
C. Colijn and E. R. Vrscay, Found. Phys. Lett. \textbf{16}, 303 (2003); quant-ph/0304198 (2003).%

\bibitem{Struyve}
W. Struyve, W. De Baere, J. De Neve, and S. De Weirdt, Phys. Lett. A \textbf{322}, 84 (2004).%



\end{thebibliography}
\end{document}